# Fine topological structure of coherent complex light created by carbon nanocomposites in LC

Vlad. V. Ponevchinsky[a], Andrey I. Goncharuk[b], Serguey S. Minenko[c], Longin N. Lisetskii, Nikolai I. Lebovka, Marat S. Soskin,

[a]Institute of Physics, NAS of Ukraine, 46 Nauki Prosp., Kiev, 03028, Ukraine, +38-044-525-55-63;
[b]Institute of Biocolloidal Chemistry, named after F. Ovcharenko, NAS of Ukraine, 42 Vernadskii Prosp., Kiev, 03142, Ukraine,+38- 044-424-03-78;
[c]Institute of Scintillation Materials of SCT "Institute for Single Crystals" of the NAS of Ukraine, 60 Lenin Ave., Kharkov, Ukraine, +38-057-341-03-58.

**ABSTRACT**

Fine complex light structure, optical singularities and electroconductivty of nematic 5CB doped by multi-walled carbon nanotubes (MWCNTs) were investigated. MWCNTs gather spontaneously to system of micro scale clusters with random fractal borders at small enough concentration. They are surrounded by the striped micro scale cladding which creates optical singularities in propagating laser beam. Applied transverse electric field above the Freedericksz initiates homeotropic arrangement of 5CB and the striped inversion walls between nanotubes clusters what diminishes free energy of a composite. Theory of their appearance and properties was built. Simultaneously the striped cladding disappears what can be treated as new mechanism of structure orientation nonlinearity in nonlinear photonics. Polarization singularities (circular C points) were measured firstly. Percolation of clusters enhances strongly electrical conductivity of the system and creates inversion walls even without applied field. Carbon nanotubes composites in LC form bridge between nano dopants and micro/macro system and are promising for applications. Elaborated protocol of singular optics inspection and characterization of LC nanocomposites is promising tool for applications in modern nanosience and technique.

## 1. INTRODUCTION

Nano science and nano technology are hot spots of modern science and technique. Very actual are carbon nanotubes as dopants of liquid crystals (LC) [1]. LC possesses simultaneously properties of both solids and liquids. They are endowed with continuous symmetries and physical prevalence of *correlations of orientations* over correlations of position [2]. Each flexible open system chooses the structure with minimal free energy. All numerous structures of LC systems obey this principle including topology of all realized and investigated 5CB carbon nanocomposites. Nematic 5CB correspond optically to the *uniaxial* crystal with high enough birefringence. It is very promising for realized firstly striped structures in LC host.

Carbon nanotubes (CNT) have *nano* scale diameter and *micro* scale length. Therefore, they possess features typical for nanoparticles which are investigated by electronic microscopes of various types. Simultaneously, their complexes and aggregates have tens and even hundreds of micrometers dimensions. Therefore, their general structure can be observed and investigated successfully by high-resolution microscopes. Managed liquid crystals elements are widely used as screens in computers, etc. Many scientific groups and companies in the world are dealing with LCs + CNTs composites and attempts of their practical applications. Optics was used up to now mainly for illustration of general structure of nano composites only. We started detailed investigation of carbon nanotubes aggregates and LC matrices fine topology by methods of singular optics [3] practically from 'clean sheet'.

The Nobel Price was given recently for realization and investigation of the graphen, first stable 2D crystal with unique parameters and properties. Graphen can roll by special technique into single-walled carbon nanotubes (SWCNT) with 1-2 nm diameters. The more mechanically stable multi-walled nanotubes (MWCNT) were subject of our investigations.

They possess few tens of nanometers outer diameters and micro scale length contrary to nano scale dimensions of used nematic 5CB. We started recently systematic investigations of fine polarization optical features of LC nanocomposites by the methods of crystal optics [3] and singular optics [4].

## 2. LC CARBON NANOCOMPOSITES

Molecules of nematic 5CB have 1.5×2.3 nm size only. They look as dog against carbon nanotubes elephant. Nevertheless, they interact strongly because two benzene rings of 5CB molecule possess practically the same dimensions as nanotubes honey-comb lattice. Therefore, their van der Waals (VDW) attraction is extra strong. Extra high anchoring energy – 2 eV is two orders higher then phonons energy at room temperature [5, 6]. Nanotubes are nailed to lateral surface of all nanotubes what influences strongly on optical and electrical parameters of 5CB carbon nanocomposite. 5CB molecules start cover nanotubes lateral surface just on the moment of LC nanocomposite preparation. So, combination of 5CB and carbon nanotubes is special one and is called "scientific duo" [6]. The 'long'/'short' MWCNTs with 5÷10/1÷2 μm length were exploit. The first our measurements have shown that nanotubes form clusters [7, 8]. The theory of CNT aggregation process was elaborated also [9]. Aggregation started from slow Brownian drift of single nanotubes. They gather then to 3D nanotubes clusters due to van der Waals attraction. This incubation process is lasted nearly one week [10]. Final clusters are stable more then one year if used LC cells are encapsulated carefully.

Visible clustering of nanotubes started at their very low concentration. Clusters are enlarged with concentration grow up to formation of the percolation structure when all neighbor cluster touch [7, 8, 11]. This moment electrical conductivity enlarged nonlinearly on three-four orders [8]. It is natural to wait that due to stochastic process of clusters formation, density of CNTs inside clusters is not fully homogeneous. Some small micro scale volumes with extended density of nanotubes have to exist. Ramified stochastic boundaries of nanotubes clusters possess fractal nature [9] what defines most of measured new optical peculiarities of LC carbon nanocomposites. The inner heterogeneities of clusters explain observed cladding heterogeneity along clusters border. A laser beam propagating through 20 μm thick LC sandwich cell diffracts and scatters few times on the fractal boundaries of clusters what created multitude of OVs already in the output plane of LC cell (Fig.1).

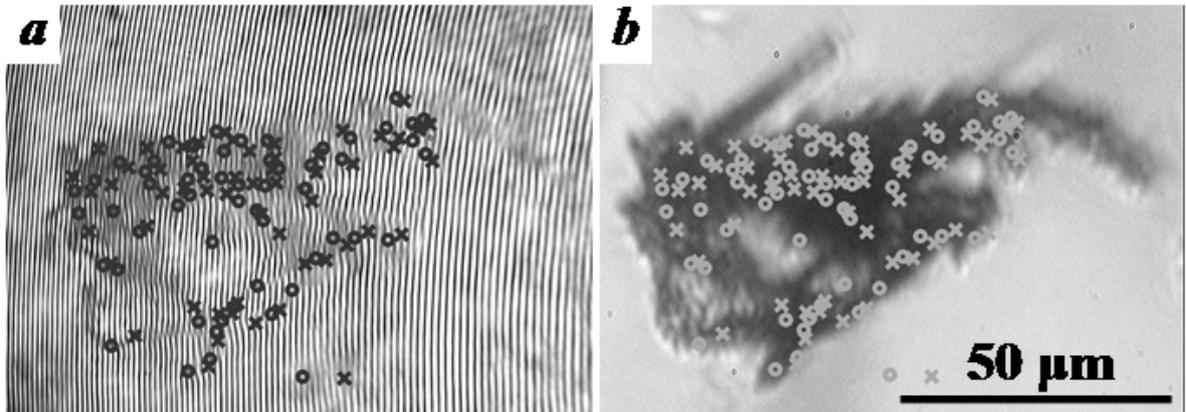

Fig.1. Optical vortices induced in propagating beam of He-Ne laser on the output plane of LC cell with CNTs: *a* –identification of optical vortices by two-beam interference with "fork patterns", *b*- cluster of nanotubes with optical vortices born along its outer and inner boundaries, x(+1), o(-1) vortex.

One of the central notions of any LC is the director **n**, which characterizes the *averaged* preferential orientation of anisometric molecules long axes [12]. But individual 5CB molecule is an anisometric one. This contradiction is resolved by formation of 5CB dimmers [13]. Typical minimal volume of LC with definite value of director possesses is ~ $(10l)^3$ where $l$ is length of LC molecule [14] ($l$=2.3 nm for 5CB host). Therefore volume with $(23)^3$ nm$^3$ is enough for definition of **n(r)**. This moment is one of the central for understanding of new fine features of LC host with nanocomposite. But fractality of stochastic borders of CNT clusters shows that orientation of nanotubes changes so quickly and randomly that director value isn't appropriate parameter for description of this inside micro scale volume between protruding ends of outer nanotubes. Contrary, it is correct for micro scale cladding out of this volume what was measured and investigated carefully.

## 3. POLARIZATION SINGULAR OPTICS OF 5CB CARBON NANOCOMPOSITES

Multitude of 5CB molecules is dismembered to three groups according their optical properties [6]: (*i*) molecules disposed far enough from nanotub*es* clusters forming LC *host, (ii)* molecules of clusters micro scale *cladding* (interfacial LC layer) around them, (*iii*) *hidden molecules* inside clusters, not observable by optical methods. Fraction of these parts depends crucially from nanotubes concentrations: *isolated* CNT clusters at their low concentrations and the *percolation* structure when neighbor clusters touch, LC host is broken on isolated 'lakes', and length of clusters boundaries enlarges drastically. The heterogeneous LC cladding started by 5CB molecules anchored on micro scale outer prominent ends of MWCNT and extends up to unperturbed LC host. In fact, some general qualitative thoughts without any greater detail were given in some previous paper [5, 14, 15]. Therefore we have observed and investigated them firstly, due to our best knowledge. LC cladding possesses micro scale width and heterogeneous striped structure even without applied field due to stochastic homeotropic orientation of 5CB molecules on boundaries of a CNT cluster and planar arrangement of 5CB host.

Very informative appeared reaction of LC carbon nanocomposites on electric field applied across LC cell below and above the Freedericksz threshold when planar oriented LC molecules are turned to homeotropic arrangement [11]. Most sensitive is the optical scheme with crossed polarizer and analyzer and LC cell in-between turned on $45°$ to involve effectively both ordinary $n_o$ and extraordinary $n_e$ refractive indices in the co-axial interference process in optimal way. It's known that alignment-inversion walls in nematics appear in the presence of magnetic and electric field [13]. We have established topological regularities of host and cladding evolution during nematic 5CB transition from planar to homeotropic arrangement. The clearly defined claddings exist without applied field (Fig.2*a*). It diminishes strongly above the Freedericksz threshold (Fig.2*b*) and disappears practically at higher voltage (Fig.2*c*).

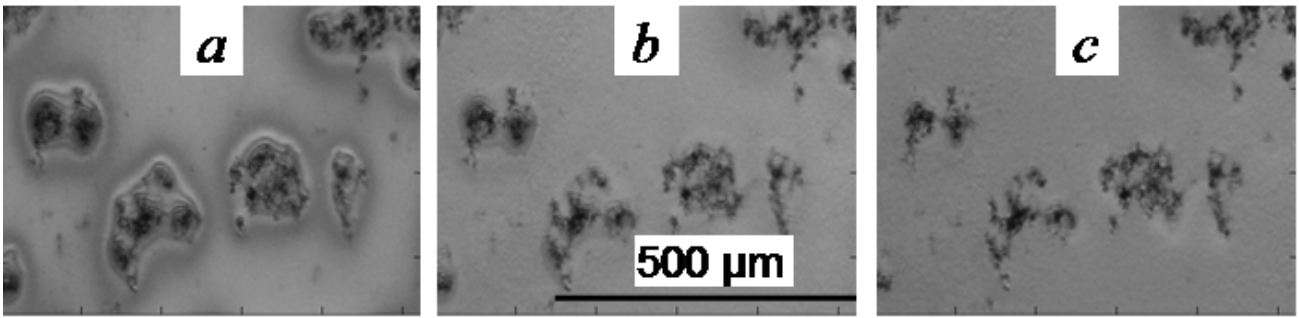

Fig.2. Cladding of 5CB around isolated nanotubes clusters without applied field (*a*), its diminishing at 1,5V exceeding the Freedericksz threshold 0,74V (*b*), and practically full disappearance at 2,5V (*c*) for "long" nanotubes at C = 0.01 wt. %.

It was observed firstly appearance of few-stripes inversion walls between bumps of neighbor clusters when electric field exceeding the Freedericksz threshold was applied (Fig.3) (see also [9, 15]). Observed inversion walls aren't smoothed domain walls existing in ferroelectrics, segnetoelectrics, because polarization state of light on opposite sites of them were the same. The reason of their birth is the essential diminishing of nanocomposite free energy under torque of LC molecules induced by applied field. Cladding is striped strongly around all clusters because descent from nearly homeotropic orientation on their inner borders to planar orientation of 5CB molecules in the host. This observed effect was measured quantitatively (see Fig.6 below).

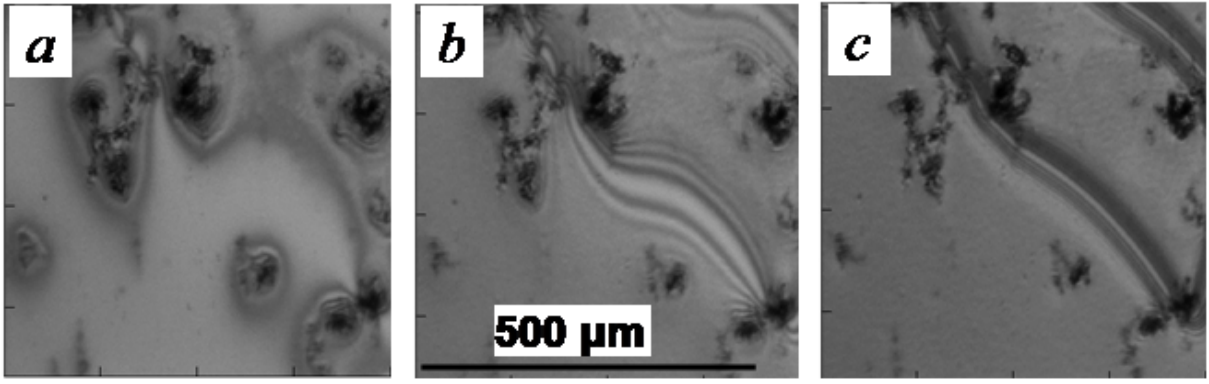

Fig.3. *a* – integrated complicated structure of claddings between near located nanotubes clusters (0.01 wt. %); *b* - birth of inversion walls between bumps of facing clusters at 1.5V, *c* – stressing of inversion walls cross-section at 2.5V.

We have developed the theory to understand origin and physics of observed inversion stripes and patterned cladding, their topological structure at various magnitude of applied field starting from pioneer theoretical analysis derived in [17]. It predicts *smooth* director tilt $\theta$ across inversion walls averaged for LC cell along z axis of beam propagation:

$$\theta(x) = \arctan\left[\exp\left(\sqrt{\frac{\varepsilon_0 \Delta\varepsilon U^2}{K d^2}}x\right)\right] \quad (1)$$

Here $\varepsilon_0$ is the dielectric constant, $\Delta\varepsilon$ – the dielectric anisotropy of used LC, $U$ is applied electric field, $d$ – thickness of LC cell, and $K$ – the averaged elastic Frank constant. For used nematic 5CB host these parameters are next: averaged Frank constants $K_{11} = K_{22} = 10^{-4}$ dyn, $\Delta\varepsilon = 13$. The LC cell with thickness 20 μm was used. Director of the uniaxial nematic is oriented always along axis its optical indicatrix [13]. Therefore, optical indicatrix is tilting together with director on the same angle. The projection of extraordinary refractive index $n_e^{eff}$ on the direction of transparency of crossed analyzer is the function of tilt angle $\theta$ also [13]:

$$n_e^{eff}(\theta) = \frac{n_o n_e}{\sqrt{n_o^2 \cos^2\theta + n_e^2 \sin^2\theta}} \quad (2)$$

Of course, it is independent from angular position of LC cell between polarizer and analyzer. As was shown before, director of planar oriented 5CB molecules is parallel to the axis of optical indicatrix. Therefore, its tilt changes smoothly the effective refractive index $n^{eff}_e(\theta)$ for the extraordinary wave going out through the crossed analyzer. Contrary, refractive index $n_o$ doesn't depend from indicatrix orientation. Both co-propagating orthogonal components of ordinary and extraordinary waves acquire their own phase changes what defines total accumulated output phase difference $\Delta\varphi(\theta)$. In reality, actual one is only phase difference above integer of $2\pi$ cycles $0 < \delta\varphi(\theta) < 2\pi$.

$$\Delta\varphi(\theta) = \frac{\left(n_e^{eff}(\theta) - n_o\right)d}{\lambda} \qquad \delta\varphi(\theta) = \Delta\varphi(\theta) - 2m\pi \quad (3)$$

The output intensity varies *periodically* with smooth growth of this output phase difference. So, we conclude that observed stripes in inversion walls are results of **interference** of co-propagating orthogonal components of host refractive index:

$$I(x) = \{1 - 2\cos(\alpha)\sin(\alpha)\cos[\delta\varphi(x)]\} \quad (4)$$

The formula (4) is written for arbitrary azimuth α of LC cell between crossed polarizer and analyzer. It's seen that stripes are absent when cell is parallel oriented to polarizer and analyzer. Change of cell angular position shifts position of stripes. When analyzer is parallel to polarizer, maxima of shown patterns became minima and vise versa. It is easy to see that used in experiment 45° orientation of LC cell between crossed polarizer and analyzer is *optimal* value of angle α. Derived analysis possesses general importance for all types of textures and

Smooth director's tilt and striped structure of inversion wall were calculated for used nematic 5CB and two applied fields above the Freedericksz threshold (Fig.4). The accumulated tilt of director equals π. It' seen that grows of applied field initiates more quick tilt of director and diminishes distance between intensity patterns. Smooth director's tilt is typical for various textures and schlieren structures in LCs [2, 18-20]. Therefore, striped intensity structures of this origin have to exist in many LC systems.

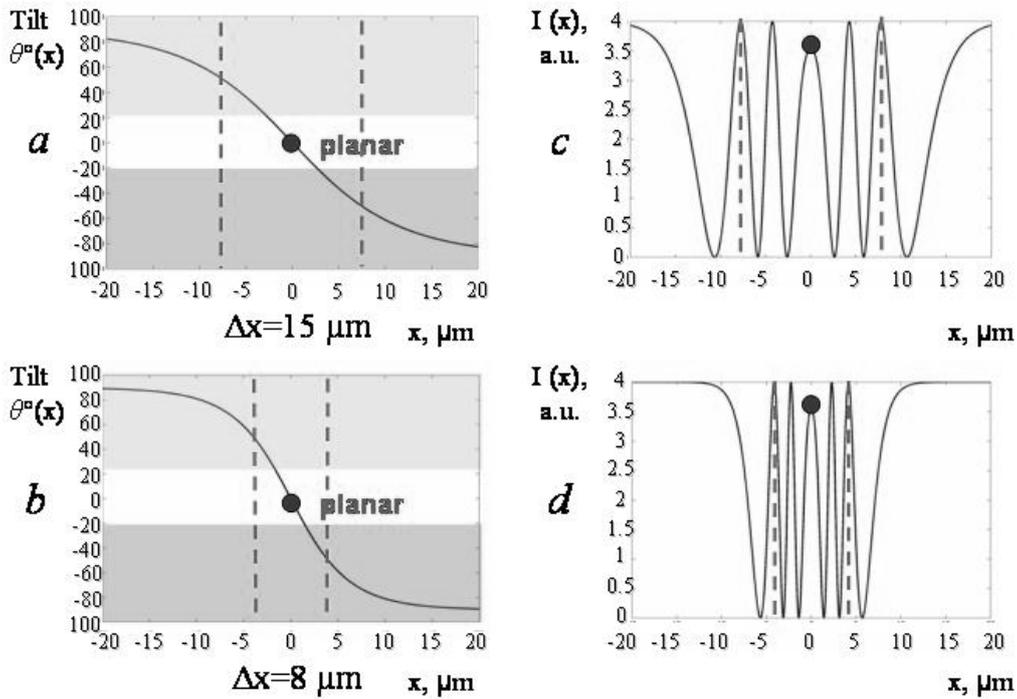

Fig.4. Calculated angles $\theta(x)$ of smooth director tilt from its initial planar arrangement (*a, b*) and intensity distribution *I*(x) across induced striped inversion walls at 0.045and 0.075 V/μm (*c, d*).

Experimental results were in full agreement with elaborated theory. Typical structure of induced inversion walls are shown in Fig.5. It's seen their width diminishes with grows of applied field.

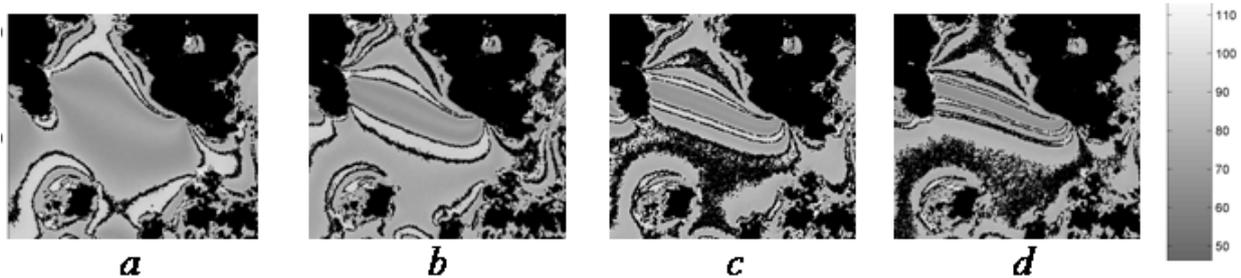

Fig.5. Stress of half-width of inversion walls in 5CB host between carbon nanoclusters with grows of applied field: *a* –1,1V, *b* – 1,2V, *c* – 1,3V, *d* – 1,5V. The gray scale shows the phase difference between ordinary and extraordinary beams, black areas stand for nanotube clusters, black points are the points with circular polarization

To obtain quantitative characteristics for cladding and inversion walls variation of polarization ellipses shape was measured by the stokes-polarimetry across them (Fig.6).

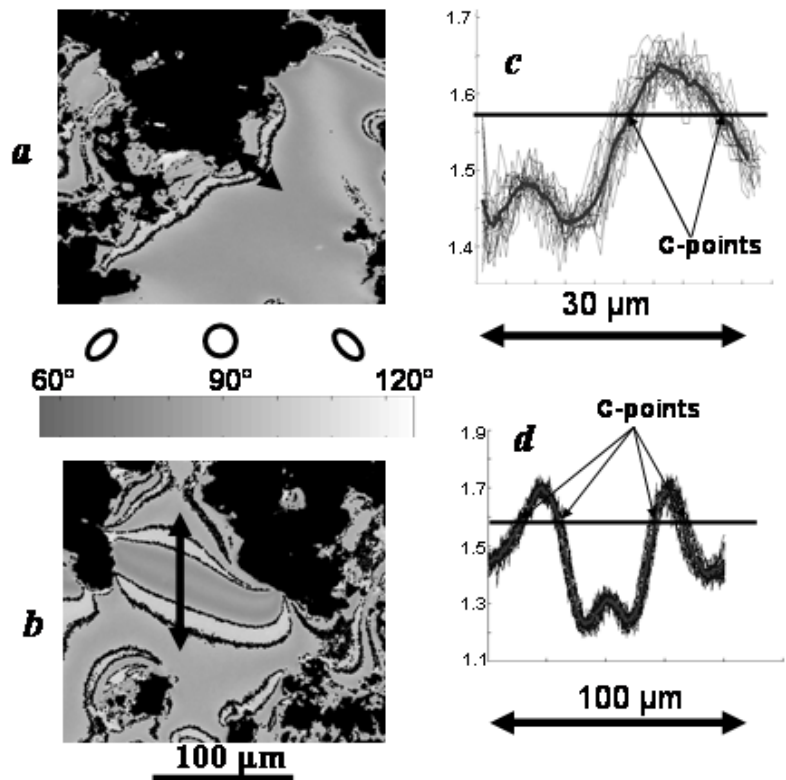

Fig.6. Polarization singular optics of 5CB carbon nanocomposites: *C* points in inversion walls and cladding cross-section at 1,2V (black areas are nontransparent nanotubes clusters): *a* – striped 5CB cladding, *b* – striped inversion walls between clusters, *c, d* – phase difference between ordinary and extraordinary beams in cross-section (black line corresponds to pure circular polarization).

The seen black points are singular C points with pure circular polarization. 5CB host without field is homogeneous enough and possesses C points on the composites border. Evolution of polarization ellipses is seen from Fig.6*c*, *d*. Each intensity extreme correspond maxima/minima of polarization ellipses shape. Two/four C points exist in cladding/inversion walls. The width of cladding is nearly twice less then full width of inversion walls. Reason of this difference is evident. 5CB molecules are oriented homeotropically in stochastic manner on the clusters border and director tilt diminishes in striped manner to planar orientation of 5CB host. Contrary, director tilt across inversion walls started from planar, grows to homeotropic orientation, diminishes to planar on the top, and diminishes smoothly trough homeotrop to planar orientation on the opposite side of inversion walls. Such type inversion walls where observed in clean 5CB cell with optical singularities ([16]). But intensity stripes were interpreted qualitatively wrong as disclinations with knee along line director tilt across inversion walls (Fig.4 in [16]), what don't correspond to real one-hill curves of director tilt [15]. It should be noted that the multi-striped inversion walls of pure LC samples (see e.g. [16-22]) could be the same origin as considered above. This will proofed in future.

When concentration of nanotubes grows up to formation of percolation structure of nanocomposites, LC host is broken on the isolated "lakes". Stresses of LC host along rough foils of clusters enlarge essentially and can overcome elasticity limit of LC host. This time striped inversion wall can appear between neighbor clusters and even bumps of a cluster without applied electric field (Fig.7*a*). Many striped inversion walls appear above Freedericksz threshold (Fig.7*b*). Their width diminishes when field grows (Fig.7*c*).

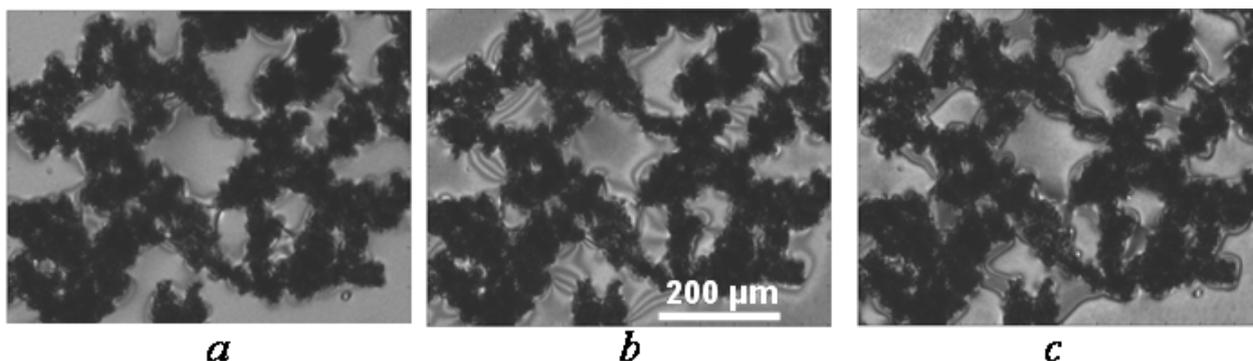

Fig.7. Inversions walls in 5CB host for percolated CNT clusters of 'long' nanotubes (C = 0.1 wt. %): *a* - without applied transverse electric field, *b* - at 1,9V, *c* - at 2,8V.

## 4. DETECTION of SUB-WAVELENGTH CNT CLUSTERS

Transition of 5CB matrix with nanocomposites to isotropic liquid in LC cell was investigated by fixation of its structure during heating of used cell (Fig.8). Many small cladding islands of micron scale dimensions were seen in LC state (Fig.8*a*) with small nanotubes aggregates and without them (Fig.8*a*). It was noticed that cladding around all clusters disappears completely (Fig.8*b*). This is natural because isotropic liquid don't possess elastic stresses in usual conditions. But the unexpected new effect was fixed. All isolated islands of cladding *without visible* nanotubes nuclei disappear (Fig.8*a, b*). This witnesses definitely that these clusters nuclei possess sub-wavelength dimensions and can't be observed by usual optical microscopes. Discover phenomenon can be treated as the new unusual specific 'super resolution effect' for LC nanocomposites. It has to be valid for all kinds of nanocomposites.

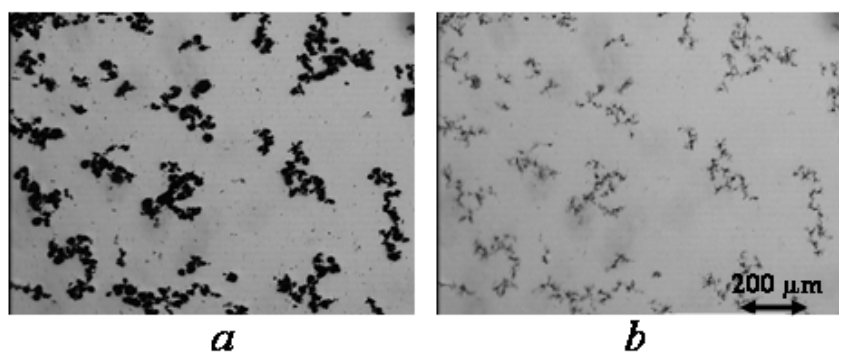

Fig.8. *a* –the typical fragment of 5CB carbon nanocomposite ("long" nanotubes, 0.025 wt. %), *b*- the same fragment after 5CB melting to isotropic liquid.

## 5. PROTOCOL of LC NANOCPOMPOSITES INSPECTION and CHARACTERIZATION by SINGULAR OPTICS TECXHNIQUE

Due to our hard opinion totality of elaborated methods are valid and promising for all kinds of LC nanocomposites. The can be dissimilated on two gropes: (*i*) general inspection of optical structure peculiarities and (*ii*) characterization of main nanocomposites parameters. They can be classified for example on next items listed below.

### 5.1. General Inspection of LC nanocomposites
The general inspection of LC nanocomposites has to be fulfilled by a modern high-resolution polarization microscope. It includes examination of visible nanoaggregates existence, their shape and dependence from nano dopants concentration.

Very informative are structure of nanoaggregates boundaries, existence of LC cladding around them and general reaction on applied field (inversion walls).

### 5.2. LC nanocomposites characterization

This set of measurements has to be done by a modern high-resolution polarization microscope equipped with rotary polarizer and Stokes analyzing unit (quarter-wave plate and analyzer) [3], interference filter with central pick of efficiency on the used laser (typically He-Ne laser) and high-resolution CCD camera.

Next optical parameters of LC nanocomposites have to be measure precisely:
- optical singularities existence, their topological charge and location,
- fractality of nanoaggregates boundaries,
- Stokes polarimetry of LC host
- Stokes polarimetry of cladding and inversion walls cross-section in actual limits of variation for applied transverse electric field.

Totality of measured LC nanocomposites characteristics will be important information of their physics and perspective of elaboration on their base command polarizers and other elements actual for tunable screens, etc. Therefore, we dare to recommend them to all numerous scientific groups and companies.

## 6. CONCLUSION

As we have shown firstly, carbon nanotubes in nematic 5CB are self-organized spontaneously to 3D micro scale clusters of random oriented nanotubes with fractal boundaries in LC nanocomposites. Strong anchoring of 5CB molecules on nanotubes side walls initiates stressed micro scale cladding around their clusters. Elliptically polarized striped inversion walls between clusters and around them are born at applied transverse electric field above the Freedericksz transition. They minimize free energy of LC nanocomposites. Polarization varies across inversion walls. The main of findings can be summarized as follows:

- Carbon nanotubes (CNT) in nematic 5CB gather spontaneously to micro scale aggregates with *fractal* borders and LC cladding. They created multitude of *optical vortices* in propagating beam;
- Strong anchoring of 5CB molecules on CNTs leads to formation of stressed micro scale striped *cladding* on borders of nanotubes clusters;
- Applied electric field creates striped *inversion walls* between CNT clusters. According to elaborated theory, intensity stripes in claddings and inversion walls appear due to *coaxial interference* of ordinary and extraordinary components of light beam. Observed striped structure has to exist in many textures and schlieren structures of LC including structures with disclinations.
- Elaborated technique of polarization singular optics is new effective tool for various LC systems *inspection* and *characterization.*
- LC nanocomposites form promising "road map" from nano world to micro/macro effective managing elements.

Due to our opinion, obtained results about LC nanocomposites are essential addition to optic and general LC physics at whole. Shown both theoretically and experimentally co-axial interference as the optical origin of inversion walls can be valid also for some inversion walls in a pure nematic and schlieren textures in LC with dopants. The same concerns demonstration of decisive influence of inner elastic stresses in nematic host on its optical structure. The elaborated protocol of polarization singular optics of LCs possesses general importance. It can be recommended for measurements and characterization of various kinds of LC cells and matrices.

## ACKNOWLEDGEMENTS

This work was supported by Projects 2.16.1.4 and 2.16.1.7 NAS of Ukraine and ITSU Project 4687.

# REFERENCES

[1] Lagerwall, J. P. F., Scalia, G., "Carbon nanotubes in liquid crystals", J. Mater. Chem. 18, 2890-2898 (2008).
[2] O.D. Lavrentovich, "Defects in liquid crystals: surface and interfacial anchoring effects" in H. Arodz et al. (eds.), "Patterns of Symmetry Breaking", 161-195 (Kluwer Academic Publishers, Printed in the Netherlands, 2003).
[3] M. Born and E. Wolf, "Principles of Optics", 7th ed. (Pergamon, New York, 1999).
[4] M. S. Soskin and M. V. Vasnetsov, "Singular optics" (E. Wolf., Progress in Optics 42,Elsevier Science B.V. ,2001) pp 219-276.
[5]] Seok Jin Jeong, Kyung Ah Park, Seok Ho Jeong et all. "Electroactive Superelongation of Carbon Nanotube Aggregates in Liquid Crystal Medium", Nano Lett. 7, 2178-2182 (2007).
[6] Rahman M., Lee W., "Scientific duo of carbon nanotubes and nematic liquid crystals", J. Phys. D: Appl. Phys. 42, 063001 (1-12) (2009).
[7] Ponevchinsky, V. V., Goncharuk, A. I., Lebovka, N. I., Soskin, M. S., "Cluster self-organization of nanotubes in nematic phase: the percolation behavior and appearance of optical singularities", JETP Lett. 91, 239-242 (2010).
[8] V. Ponevchinsky. A. I. Goncharuk, V. I. Vasil'ev, N. I. Lebovka, M. S. Soskin, "Optical singularities induced in a nematic-cell by carbon nanotubes", SPIE 7613, 761306 (10 pp.) (2010).
[9] V. V. Ponevchinsky, A. I. Goncharuk, S. S. Minenko, L. N. Lisetski, N. I. Lebovka, M. S. Soskin, "Incubation Processes in Nematic 5CB + Multi-Walled Carbon Nanotubes Composites: Optical Singularities and Inversion Walls, Percolation Phenomena", Nonlin. Optics, Quant. Optics (in press).
[10] L. N. Lisetski, A. M. Chepikov, S. S. Minenko, N. I. Lebovka, M. S. Soskin, "Dispersion of carbon nanotubes in nematic liquid crystals: effect of nanotubes geometry", Functional Materials 18, 143-149 (2011).
[11] L. N. Lisetski, S. S. Minenko, V. V. Ponevchinsky, M. S. Soskin, A. I. Goncharuk, N. I. Lebovka, "Microstructure and incubation processes in composite liquid crystalline material (5CB) filled by multi-walled carbon nanotubes", Mat. Sci. Eng. Technol. 42, 5 (2011).
[12] Vlad. V. Ponevchinsky, Andrey.I. Goncharuk, Sergei V. Naydenov, Longin N. Lisetski, Nikolai I. Lebovka, Marat S. Soskin, "Complex light with Optical singularities induced by nanocomposites", SPIE 7950, 79500A (10 pp.) (2011).
[13] L.M. Blinov, "Structure and Properties of Liquid Crystals" (Springer Science + Media B.V. 2011).
[14] Prof. V.Yu.Reshetnyak, private communication (2011).
[15] Giuy Scalia, Jan P. F. Lagerwall, M. Haluska et al., "Effect of phenyl rings in liquid crystal molecules on SWCNTs studied by Raman spectroscopy", Phys. Stat. Sol. (b) 243, 3238-3241 (2006).
[16] R. Basu and G. S. Iannacchione, "Orientational coupling enhancement in a carbon nanotube dispersed liquid crystal" Phys. Rev. E 81, 051705 (2010).
[17] W. Heilfrich, "Alignment-inversion walls in nematic liquid crystals in the presence of a magnetic field", Phys. Rev. Lett. 21, 1518-1521 (1968).
[18] P. G. de Genne and J. Prost, The Physics of Liquid Crystalls, 2nd ed. (Oxford Univ., Oxford, 1995).
[19] S. Chandrasekhar, Liquid crystals, (Cambridge Univ. Press, 1977).
[20] Demus D. and Richter' L., Textures in Liquid Crystal (Ferlag Chemie, Weinheim, N. Y., 1978).
[21] A. Nesrullajev, "Peculiarities of inversion walls and singular points in oriented textures of nematic mesophase", Cryst. Res. Technol. 44, 747-753 (2009).
[22] P. E. Cladis, W. van Saarloos, P. L. Finn, and A. R. Kortan, "Dynamics of Line defects in Nematic Liquid Crystals", Phys. Rev. Lett. 58, 222-225 (1987).
[23] J. M. Jilli, S. Thiberge, A. Vierheilig, and F. Fried, „Inversion walls in homeotropic nematic and cholesteric layers", Liq. Cryst. 23, 619-628 (1997).
[24] I. Jánossy and S. K. Procad, "Optical generation of inversion walls in nematic liquid crystals", Phys. Rev. E 63, 041705 (7 pp.) (2001).
[25] Tomoyuki Nagaya, Jean-Mark Gilli, "Experimental Study of Spinodal Decomposition in 1D conserved Order parameter System", Phys. Rev. Lett. 92, 145504 (4 pp.) (2004).